\documentstyle[12pt]{article}

\begin{document}
\pagestyle{empty}

\begin{flushright}
{CERN-TH.7208/94}
\end{flushright}

\vspace*{5mm}

\begin{center}
{\large \bf The  next-to-leading QCD approximation \\
to the Ellis-Jaffe sum rule } \\
\vspace*{1cm}
{\bf S.A. Larin}
\footnote{Permanent address: INR,  Moscow 117312, Russia.} \\
\vspace*{0.3cm}
Theory Division, CERN, CH - 1211, Geneva 23, Switzerland\\
\vspace*{2cm}
\end{center}

\vspace*{5mm}

\begin{abstract}
The $\alpha_s^2$ correction to the Ellis-Jaffe
sum rule for the structure function $g_1$ of polarized deep inelastic
lepton-nucleon
scattering is calculated.
\end{abstract}

\vspace*{5cm}

\begin{flushleft} CERN-TH.7208/94 \\
March 1994
\end{flushleft}

\addtocounter{page}{-1}
\thispagestyle{empty}
\vfill\eject
\pagestyle{empty}
\pagestyle{plain}
\setcounter{page}{1}

\newpage
The results of the EMC Collaboration at CERN \cite{emc} and E130 Collaboration
at SLAC \cite{e130} for  the Ellis-Jaffe \cite{ej} sum rule
$\int_{0}^{1} dx g_1^p(x,Q^2)$ attracted a lot of attention to this sum rule;
see \cite{et}-\cite{anr}
and references therein. Recent data of the SMC Collaboration at CERN \cite{smc}
 on
polarized scattering of muons off deuterium
and of the E142 Collaboration at SLAC \cite{e142}
on polarized scattering of electrons off helium $^3He$ allowed the
determination of
the analogous sum rule  $\int_{0}^{1} dx g_1^n(x,Q^2)$ for a neutron. This
in turn allowed us to find a
difference $\int_{0}^{1} dx [g_1^p(x,Q^2)-g_1^n(x,Q^2)]$ which is the
Bjorken sum rule \cite{b}.
At present, the Bjorken sum rule is calculated within QCD with quite high
accuracy.
The
$\alpha_s$ correction \cite{kmmus}, $\alpha_s^2$ correction \cite{gl} and
$\alpha_s^3$
correction \cite{lv} are calculated in the leading twist approximation.
The higher twist
corrections are also calculated \cite{bk}.
For the Ellis-Jaffe sum rule only the $\alpha_s$ correction  was calculated
in the leading twist \cite{k}. The power corrections were calculated
in \cite{bbk}.

In the present paper we obtain the $\alpha_s^2$ correction to the Ellis-Jaffe
sum rule
in the leading twist massless quark approximation.
All calculations are performed in dimensional regularization \cite{hv}.
Renormalizations
are done within the $\overline{MS}$-scheme \cite{bbdm}, the standard
modification
of the Minimal Subtraction scheme \cite{h}.

Polarized deep inelastic electron-nucleon scattering is described by
the hadronic tensor
\[
W_{\mu\nu}(p,q)  = \frac{1}{4\pi} \int d^4 z \  e^{iqz}
<p,s\mid J_\mu(z) J_\nu(0)
\mid p,s> =
\]
\[= \left(-g_{\mu\nu} + \frac{q_\mu q_\nu}{q^2}\right) F_1(x,Q^2)
     +\left(p_\mu-\frac{p\cdot q}{q^2} q_\mu\right)
     \left(p_\nu-\frac{p\cdot q}{q^2}
      q_\nu\right)\frac{1}{p\cdot q} F_2(x,Q^2)+
\]
\begin{equation}
\label{hadtensor}
+i \varepsilon_{\mu\nu\rho\sigma} q_\rho
	\left[ \frac{s_\sigma}{p\cdot q} g_1(x,Q^2)
	+ \frac{s_\sigma p \cdot q - p_\sigma q \cdot s}{(p \cdot q)^2}
	g_2(x,Q^2) \right],
\end{equation}
of which we will consider the structure function $g_1$.
Here $J_{\mu}= \overline{\psi} \gamma_\mu \hat{E} \psi
=\sum_{i=1}^{n_f}e_i\overline{\psi}_i \gamma_\mu  \psi_i$
is the electromagnetic quark current
and $\hat{E}=diag(2/3,-1/3,-1/3,...)$ is the quark
electromagnetic charge matrix.
$x = \frac{Q^2}{2 p \cdot q}$ is the Bjorken variable, $Q^2 = -q^2$.
 The nucleon state $\mid p ,s>$  is covariantly normalized as
$<p,s\mid p',s'>=\delta_{ss'}2p^0(2\pi)^3\delta^{(3)}({\bf p}-{\bf p'})$.
$s_{\sigma}$ is the polarization vector of the nucleon:
$s_{\sigma}=\overline{U}(p,s)\gamma_{\sigma}\gamma_5U(p,s)$, where
$U$ is the nucleon spinor, $ \overline{U}(p,s)U(p,s)=2M$.

The moments of the deep inelastic structure functions are expressed \cite{chm}
via quantities of the Wilson operator product expansion (OPE) of the
corresponding currents.
We need the OPE of two electromagnetic currents.
The strict method of the OPE ensures \cite{c},\cite{chs} that
the OPE of two gauge-invariant currents can contain only
gauge-invariant operators with their renormalization basis.
Thus we have only contributions from the non-singlet and singlet axial
currents in
the OPE of electromagnetic currents
in the leading twist for the considered structure:
\[	i \int dz \  e^{iqz} T\{ J_\mu(z) J_\nu(0) \}
		\stackrel{Q^2 \rightarrow \infty}{=}
\]
\begin{equation}
\label{ope}
		=\varepsilon_{\mu\nu\rho\sigma} \frac{q_\rho}{q^2}\left[
\sum_{a}C^a(\log(\frac{\mu^2}{Q^2}),a_s(\mu^2) )J_{\sigma}^{5,a}(0)
+C^s( \log(\frac{\mu^2}{Q^2}),a_s(\mu^2) )J_{\sigma}^{5}(0)
+higher~ twists \right],
\end{equation}
where the non-singlet contribution can be rewritten as
\[
\sum_{a}C^aJ_{\sigma}^{5,a}(0)=
C^{ns}\sum_{a}Tr(\hat{E}^2t^a)J_{\sigma}^{5,a}(0),
\]
to introduce, as usual,
the unique non-singlet coefficient function $C^{ns}$ not
depending on the number $a$.\\
Here $J^{5,a}_\sigma(x) = \overline\psi(x) \gamma_\sigma
\gamma_5 t^a \psi(x)$ is
the non-singlet axial current,
where $t^a$ is a generator of a flavour group,
$Tr(t^at^b)=\frac{1}{2}\delta^{ab}$,
and $J^{5}_\sigma(x) =\sum_{i=1}^{n_f}
\overline{\psi}_i(x) \gamma_\sigma \gamma_5  \psi_i(x)$
is the singlet axial current.
The known twist-two and spin-one axial gluon current
$K_\sigma =4 \varepsilon_{\sigma\nu_1\nu_2\nu_3}(A_{\nu_1}^a\partial_{\nu_2}
A_{\nu_3}^a+\frac{1}{3}gf^{abc}A_{\nu_1}^a
A_{\nu_2}^b A_{\nu_3}^c)$ has also the necessary quantum numbers,
but it cannot contribute to
the above operator product expansion because it is not gauge invariant.
Here and further on (before presenting the final results),
we use the most practical definition for the strong coupling constant
from the calculational point of view

\[
 a_s=\frac{g^2}{16\pi^2}=\frac{\alpha_s}{4\pi}.
\]

The Ellis-Jaffe sum rule is expressed as
\[
\int_{0}^{1} dx g_1^{p(n)}(x,Q^2) = C^{ns}(1,a_s(Q^2))
\left(\pm \frac{1}{12}|g_A|+\frac{1}{36}a_8\right)+
\]
\begin{equation}
\label{ej}
+C^s( 1,a_s(Q^2) )
\exp\left(\int_{a_s(\mu^2)}^{a_s(Q^2)}da'_s\,
\frac{\gamma^s(a'_s)}{\beta(a'_s)}\right)\frac{1}{9} \Sigma(\mu^2),
\end{equation}
where some comments are in order.
Here $p(n)$ denotes a target: proton (or neutron).
The plus (minus) before $|g_A|$ corresponds to the proton (neutron) target.
The proton matrix elements of the axial currents are defined as follows:
\[
|g_A | s_{\sigma}=2<p,s\mid J_{\sigma}^{5,3}\mid p,s>
=(\Delta u-\Delta d)s_{\sigma},
\]
where $g_A/g_V=-1.2573\pm0.0028 $ \cite{pd} is the constant
of the neutron beta-decay;
\[
a_8s_{\sigma}= 2 \sqrt {3}<p,s\mid J_{\sigma}^{5,8} \mid p,s>=
(\Delta u+\Delta d-2\Delta s)s_{\sigma},
\]
\[
 \Sigma(\mu^2)s_{\sigma}=<p,s\mid J_{\sigma}^{5}\mid p,s>=
(\Delta u+\Delta d+\Delta s)s_{\sigma},
\]
and we use the standard notation
\[
 \Delta q(\mu^2)s_{\sigma}=
<p,s\mid\overline{q} \gamma_\sigma \gamma_5 q \mid p,s>,~~q=u,d,s.
\]
We omitted the contributions of the nucleon matrix elements for quarks
heavier than the s-quark but it is straightforward to include them.
We should stress here that $g_A$  and $a_8$ do not depend on the
renormalization point
$\mu^2$ since the corresponding non-singlet currents
$J_{\sigma}^{5,3}$ and $J_{\sigma}^{5,8}$ are conserved in the massless limit,
and hence their
renormalization constants are equal to one. On the contrary,
the singlet axial current has
a non-trivial renormalization constant. Hence the quantity
$\Sigma(\mu^2)$ does depend
on the renormalization point (that is why it is not a physical quantity).
The coefficient functions $C^{ns}(1,a_s(Q^2))$ and
$C^s(1,a_s(Q^2))$ are normalized in the standard way to the unity at the
tree level. The renormalization group technique was applied to the
coefficient functions
to kill logarithms $\log(\frac{\mu^2}{Q^2})$.

The singlet axial current has the non-zero anomalous dimension
$\gamma^s(a_s)$ due to the axial anomaly \cite{adler,bj}. So one can say that
the axial anomaly contributes to this sum rule through the renormalization
group exponent of the singlet current contribution in eq.(\ref{ej}).

Let us define the functions $\gamma^s(a_s)$
and $\beta (a_s)$  in the  renormalization group exponent of eq.(\ref{ej}).
The renormalization group QCD $\beta$-function is calculated \cite{tvz,lvb}
in the $\overline{MS}$-scheme at the 3-loop level :
\[
\beta (a_s)  =  \mu^2 \frac{d a_s}{d \mu^2}
           = - \beta_0 a_s^2-- \beta_1 a_s^3-- \beta_2 a_s^4
\]
\begin{equation}
\label{beta}
           = -\left(11-\frac{2}{3}n_f\right)a_s^2
                -\left(102-\frac{38}{3}n_f\right)a_s^3
             -\left(\frac{2857}{2}-\frac{5033}{18}n_f+\frac{325}{54}n_f^2
                 \right)a_s^4.
\end{equation}
To define the anomalous dimension of the singlet axial current
we should first define the singlet axial current itself within dimensional
regularization. To define the singlet axial current we will follow the
lines of ref.
\cite{l}, where the 't Hooft -Veltman definition \cite{hv} of the
$\gamma_5$-matrix
is elaborated for the multiloop case.
The singlet current $J_{\sigma}^5$ is renormalized
multiplicatively and is expressed via
the bare one $[J_{\sigma}^5]_B$ as
\begin{equation}
J_{\sigma}^5=Z_5^sZ_{MS}^s[J_{\sigma}^5]_B.
\end{equation}
Here $Z_{MS}^s$ is the $\overline{MS}$ renormalization constant which contains
only poles in the regularization parameter $\epsilon$,
the dimension of the space-time
being $D=4-2\epsilon$.
The extra finite renormalization constant $Z_5^s$ is introduced
to keep the exact 1-loop
Adler-Bardeen form \cite{ab} for the operator anomaly equation
within dimensional regularization in all orders in $a_s$:

\begin{equation}
\label{anom}
\partial_\mu J^5_\mu=a_s\frac{n_f}{2}(G\tilde{G}),
\end{equation}
where all quantities are renormalized ones.
$G\tilde{G}=\varepsilon_{\mu\nu\lambda\rho}G_{\mu\nu}^aG_{\lambda\rho}^a$
and
$G_{\mu\nu}^a=\partial_\mu A_\nu^a-\partial_\nu A_\mu^a+g f^{abc}A_\mu^b
A_\nu^c$ is the gluonic field strength tensor.

In fact the full physical quantity, the Ellis-Jaffe sum rule, does not
depend on the choice
of the normalization constant $Z_5^s$. But to be definite we adopt  the
normalization of \cite{l} to keep in the
$\overline{MS}$-scheme the singlet axial current
satisfying eq.(\ref{anom}).

The anomalous dimension of the singlet axial current
is zero at the 1-loop level and
starts from the 2-loop level.
To have the next-to-leading approximation we need two
non-zero terms, i.e. the 3-loop approximation.
The 3-loop approximation for the anomalous
dimension of the singlet axial current
was calculated in \cite{l} and confirmed in \cite{ck}.
The result in the adopted normalization reads
\[
\gamma^s(a_s) = \mu^2\frac{d\log (Z_5^sZ_{MS}^s)}{d\mu^2}
=\gamma^{(0)}a_s+\gamma^{(1)}a_s^2+\gamma^{(2)}a_s^3=
\]
\begin{equation}
\label{gamma}
       = a_s^2  (  - 6C_Fn_f )
       + a_s^3  \left[ \left(18C_F^2 - \frac{142}{3}C_FC_A\right)n_f
 + \frac{4}{3}C_Fn_f^2  \right].
\end{equation}
Here $C_{F} = \frac{4}{3}$ and $C_{A}= 3$ are the Casimir operators of the
fundamental and adjoint representation of the colour group $SU(3)$.
The $a_s^2$ term agrees with the calculation
\cite{k} after multiplication of our result
by the factor $(-2)$ due to different normalizations.

The non-singlet coefficient function $C^{ns}(1,a_s(Q^2))$ was calculated in the
$a_s^3$ approximation in \cite{lv} where the Bjorken sum rule was calculated in
this approximation.
We want to obtain the $a_s^2$ correction to the singlet contribution to the
Ellis-Jaffe sum rule (\ref{ej}):
\[
C^s( 1,a_s(Q^2) )
\exp\left(\int_{a_s(\mu^2)}^{a_s(Q^2)}da'_s\,
\frac{\gamma^s(a'_s)}{\beta(a'_s)}\right)\frac{1}{9}\Sigma(\mu^2)=
\]
\begin{equation}
=C^s( 1,a_s(Q^2) )
\left[ 1- a_s(Q^2)\frac{\gamma^{(1)}}{\beta_0}
+a_s(Q^2)^2\frac{\gamma^{(1)}\beta_1+
(\gamma^{(1)})^2 -\gamma^{(2)}\beta_0}{2\beta_0^2}\right]
 \frac{1}{9} \Sigma_{inv},
\end{equation}
where we introduced the notation
\begin{equation}
\Sigma_{inv} \equiv \exp\left(-\int^{a_s(\mu^2)}da'_s\,
\frac{\gamma^s(a'_s)}{\beta(a'_s)}\right) \Sigma(\mu^2)
\end{equation}
for the renormalization group-invariant
(i.e. $\mu^2$-independent) nucleon matrix element of the singlet axial current.

Beside the 3-loop (the order $a_s^3$)
approximation of the anomalous dimension, we need also
the 2-loop (the order $a_s^2$) approximation
for the singlet coefficient function $C^{s}(1,a_s)$
which has already  been calculated in \cite{zn}.
Here we present the calculation of
$C^{s}$ with another method to confirm the validity of  the result obtained in
\cite{zn}. We use the "method of projectors" \cite{glt}.
To project out the coefficient function  $C^{s}$ from the OPE
of eq.(\ref{ope}), one should sandwich this equation between quark states
and nullify the quark momentum $p$. To be more precise, one should consider the
following Green function
\[
	i \int dz \  e^{iqz} <0\mid
 T{\overline{\psi}(p)\gamma_{\sigma}\gamma_5\psi(p)
 J_\mu(z) J_\nu(0) } \mid0> \mid^{amputated}_{p=0}=
\]
\begin{equation}
\label{method}
		=\varepsilon_{\mu\nu\rho\sigma} \frac{q_{\rho}}{q^2}
C^s( \log(\frac{\mu^2}{Q^2}),a_s(\mu^2) )
<o\mid T{\overline{\psi}(p)\gamma_{\sigma}\gamma_5\psi(p)
Z_5^sZ_{MS}^s[J_{\sigma}^{5}(0))]_B } \mid0> \mid ^{amputated}_{p=0},
\end{equation}
where some remarks are in order. $\psi(p)$ is the Fourier transform
of the quark
field carrying the momentum $p$. Quark legs are amputated.
The essence of the method \cite{glt} is the nullification of
the quark momentum $p$.
In the dimensional regularization scheme all massless
vacuum diagrams are equal to zero. So on the r.h.s.
only the tree graphs survive after the nullification of $p$.
In our case the only operator which produces a tree graph is
$J_{\sigma}^5$. The non-singlet axial current does not contribute
because of the
nullification of the flavour trace: $Tr(t^a)=0$. It is interesting
to note that at $p=0$ we have infrared divergences
in the diagrams of the l.h.s. of eq.(\ref{method}).
But these divergences are cancelled by
the ultraviolet poles of the renormalization constant
$Z_{MS}^s$ of the singlet current.

Thus to calculate $C^s$ we need to calculate
the diagrams contributing to the l.h.s.
of eq.(\ref{method}). These are the diagrams of the
forward scattering of a photon off quarks with  photon momentum
$q$  and zero quark momentum.
In comparison with the calculation of the non-singlet
coefficient function $C^{ns}$ \cite{gl,lv},
we have at the 2-loop level two extra diagrams where
both electromagnetic vertices are inside a closed quark loop.
The analytic calculation of the diagrams has been done with the symbolic
manipulation program FORM \cite{form} by means of the package MINCER
\cite{mincer}. This package is based on algorithms of ref. \cite{t}.
The result is
\begin{equation}
\label{c}
C^s(1,a_s(Q^2))=1+a_s\left( -3C_F \right)+
a_s^2 \left[ \frac{21}{2}C_F^2-23C_FC_A
+\left( 8\zeta_3+\frac{13}{3} \right) C_Fn_f \right],
\end{equation}
where $\zeta_3$ is the Riemann zeta-function ($\zeta_3 =1.202056903\dots$).
This result of ours agrees with the calculation in \cite{zn}
if one takes into account the fact
that another finite constant $Z_5^{ns}$ (relevant for
the non-singlet axial current, see \cite{lv,l}) was taken in \cite{zn}
for the normalization of the
singlet axial current instead of our $Z_5^s$. Multiplying our result (\ref{c})
by the factor
\[\frac{Z_5^s}{Z_5^{ns}}=\frac{1+ a _s (  - 4C_F )
       + a_s^2 ( 22C_F^2 - \frac{107}{9}C_AC_F +\frac{31}{18}C_Fn_f )}
{1+ a  (  - 4C_F )
       + a^2  (  22C_F^2 - \frac{107}{9}C_FC_A + \frac{2}{9}C_Fn_f  )}+O(a_s^3)
\]
one can reproduce the result of \cite{zn}. So we have a strong check of $C^s$.

In principle our technique allows us to compute
also the $\alpha_s^3$ correction to the singlet coefficient function $C^s$.
But we
need then also the 4-loop  anomalous dimension
of the singlet axial current in order to
take into account this correction
self-consistently in the next-next-to-leading approximation
for the the Ellis-Jaffe sum rule. The calculation of the 4-loop
anomalous dimension is very time-consuming at present,
although all the necessary techniques
are available.

Collecting together all the relevant results for the coefficient functions and
the anomalous dimension, we obtain finally
the next-to-leading approximation for the Ellis-Jaffe sum rule:
\[
\int_{0}^{1} dx g_1^{p(n)}(x,Q^2)=
 \Biggl\{1 - \left(\frac{\alpha_s(Q^2)}{\pi}\right)
	+ \left(\frac{\alpha_s(Q^2)}{\pi}\right)^2
 \left( -\frac{55}{12} + \frac{1}{3} n_f \right)+
\]
\[
+ \left(\frac{\alpha_s(Q^2)}{\pi}\right)^3
 \left[ -\frac{13841}{216} - \frac{44}{9}
	\zeta_3 + \frac{55}{2} \zeta_5 +  \left( \frac{10339}{1296}
	+\frac{61}{54} \zeta_3 - \frac{5}{3} \zeta_5 \right) n_f
        - \frac{115}{648}
	n_f^2 \right] \Biggr\}\times
\]
\[
\times \left(\pm \frac{1}{12}|g_A|+\frac{1}{36}a_8\right)+
\]
\[
+\left\{1 - \left(\frac{\alpha_s(Q^2)}{\pi}\right)
	+ \left(\frac{\alpha_s(Q^2)}{\pi}\right)^2 \left[ -\frac{55}{12}
+ n_f \left(\frac{13}{36}+\frac{2}{3}\zeta_3\right)\right]\right\}\times
\]
\[
\times\left[1+\left(\frac{\alpha_s(Q^2)}{\pi}\right)\frac{6n_f}{(33-2n_f)}
+\left(\frac{\alpha_s(Q^2)}{\pi}\right)^2
\frac{\frac{1029}{4}n_f+\frac{23}{2}n_f^2+\frac{1}{3}n_f^3}{(33-2n_f)^2}
\right]\times
\]
\begin{equation}
\label{final}
\times
\exp\left(-\int^{a_s(\mu^2)}da'_s\,
\frac{\gamma^s(a'_s)}{\beta(a'_s)}\right)\frac{1}{9} \Sigma(\mu^2).
\end{equation}
Here we use $\frac{\alpha_s}{\pi}=\frac{g^2}{4\pi^2}$
for the strong coupling constant. We keep the known
(extra for the next-to-leading approximation)
$\alpha_s^3$ term \cite{lv} for the
non-singlet part, since this part determines the Bjorken sum rule.
For the singlet part we factorize the $Q^2$-dependent factors
coming from the coefficient function
(the first factor in the singlet part) and from the
renormalization group exponent (the
second factor).
The leading $\alpha_s$ term agrees with \cite{k}.

For the case $n_f=3$ the sum rule reads
\[
\int_{0}^{1} dx g_1^{p(n)}(x,Q^2)=
\left[ 1 - \left(\frac{\alpha_s(Q^2)}{\pi}\right)
	-3.5833 \left(\frac{\alpha_s(Q^2)}{\pi}\right)^2
-20.2153 \left(\frac{\alpha_s(Q^2)}{\pi}\right)^3 \right] \times
\]
\[
\times\left(\pm\frac{1}{12}|g_A|+\frac{1}{36}a_8\right)+
\]
\[
+\left[1 - \left(\frac{\alpha_s(Q^2)}{\pi}\right)
	-1.0959 \left(\frac{\alpha_s(Q^2)}{\pi}\right)^2 \right]
\left[1+0.6666\left(\frac{\alpha_s(Q^2)}{\pi}\right)
+1.2130\left(\frac{\alpha_s(Q^2)}{\pi}\right)^2\right]\times
\]
\begin{equation}
\label{final3}
\times
\exp\left(-\int^{a_s(\mu^2)}da'_s\,
\frac{\gamma^s(a'_s)}{\beta(a'_s)}\right)\frac{1}{9}\Sigma(\mu^2).
\end{equation}

The Ellis-Jaffe sum rule
looks most compact for the choice $\mu^2=Q^2$ since the renormalization
group exponent becomes a unity.
But then the $Q^2$-dependence
jumps inside the matrix element  of the singlet axial current $\Sigma$:
\[
\int_{0}^{1} dx g_1^{p(n)}(x,Q^2)=
\left[ 1 - \left(\frac{\alpha_s(Q^2)}{\pi}\right)
	-3.5833 \left(\frac{\alpha_s(Q^2)}{\pi}\right)^2
-20.2153 \left(\frac{\alpha_s(Q^2)}{\pi}\right)^3 \right] \times
\]
\[
\times\left(\pm \frac{1}{12}|g_A|+\frac{1}{36}a_8\right)+
\]
\begin{equation}
+\left[1 - \left(\frac{\alpha_s(Q^2)}{\pi}\right)
	-1.0959 \left(\frac{\alpha_s(Q^2)}{\pi}\right)^2 \right]
 \frac{1}{9} \Sigma(Q^2).
\end{equation}

The conclusion is that the $\alpha_s^2$ correction to the Ellis-Jaffe sum rule
is quite small in the $\overline{MS}$-scheme.

{{\bf Acknowledgements}}\\
I am grateful to K.G. Chetyrkin, J. Ellis,
B.L. Ioffe, M. Karliner, W.L. van Neerven
and J.A.M. Vermaseren for helpful discussions.
I would like to thank the Theory Division of CERN for warm hospitality.
The work is supported in part by the Russian Fund of the Fundamental Research,
Grant N 94-02-04548-a.

\end{document}